\documentclass[a4paper, journal]{IEEEtran}      

\renewcommand{\baselinestretch}{0.936} 

\newcommand{\myfigWidth}{0.96\columnwidth}
\newcommand*{\affmark}[1][*]{\textsuperscript{#1}}
\usepackage{cite} 
\usepackage[inkscapelatex=false]{svg}
\usepackage{graphicx}
\usepackage{amsmath}
\usepackage{caption}
\usepackage{subfig}
\usepackage{flushend} 
\usepackage{array}

\begin{document}
\title{Atrial Fibrillation Detection Using Weight-Pruned, Log-Quantised Convolutional Neural Networks}

\author{Xiu Qi Chang\affmark[1],
        Ann Feng Chew\affmark[1],
        Benjamin Chen Ming Choong\affmark[1],
        Shuhui Wang\affmark[1],
        Rui Han\affmark[2],\\
        Wang He\affmark[2],
        Li Xiaolin\affmark[2],
        Rajesh C. Panicker\affmark[1],
        Deepu John\affmark[2]
\thanks{This work is supported by Microelectronic Circuit Centre Ireland and Irish Research Council.}
\thanks{\affmark[1] X. Q. Chang, A. F. Chew, B. C. M. Choong, S. Wang and R. C. Panicker are with Department of Electrical and Computer Engineering, National University of Singapore, Singapore,
Email: \{e0203084, chewannfeng, benjamin.choong, e0202983\}@u.nus.edu,
rajesh@nus.edu.sg.
}
\thanks{\affmark[2] D. John, R. Han, W. He and  L. Xiaolin are with School of Electrical and Electronic Engineering, University College Dublin, Ireland,
Email: deepu.john@ucd.ie, \{rui.han , he.wang, xiaolin.li\}@ucdconnect.ie.}
}

\maketitle
\thispagestyle{empty}
\pagestyle{empty}

\begin{abstract}

Deep neural networks (DNN) are a promising tool in medical applications. However, the implementation of complex DNNs on battery-powered devices is challenging due to high energy costs for communication. In this work, a convolutional neural network model is developed for detecting atrial fibrillation from electrocardiogram (ECG) signals. The model demonstrates high performance despite being trained on limited, variable-length input data. \emph{Weight pruning} and \emph{logarithmic quantisation} are combined to introduce sparsity and reduce model size, which can be exploited for reduced data movement and lower computational complexity. The final model achieved a 91.1× model compression ratio while maintaining high model accuracy of 91.7\% and less than 1\% loss.
\end{abstract}


\IEEEpeerreviewmaketitle

\section{\label{Introduction}Introduction}
%
%
%
%
Deep neural networks (DNN) have proven to be an effective tool in smart healthcare, demonstrating remarkable performance in analysing and performing classification on complex data. One promising application of DNNs is the detection of cardiac arrhythmia from ECG signals, in which DNNs can provide accurate predictions from raw signals without extensive preprocessing and feature extraction~\cite{hannun2019cardiologist,xiaolin_icecs20,maryam_tbiocas2021,ANNet}. Atrial fibrillation (AFib) is a form of cardiac arrhythmia associated with coronary artery disease, heart failure, stroke, etc.~\cite{colloca2013mobile}. A robust DNN implemented in an ECG wearable can serve as a real-time detector to alert someone experiencing AFib to obtain medical assistance promptly.

However, DNN inference is memory intensive and incurs significant energy costs for data transfer and processing~\cite{zhang2020iot}, presenting a challenge for implementation on a battery-powered wearable. In particular, off-chip communication in such systems (loading from and storing to DRAM) typically consume more energy than computation~\cite{chen2016eyeriss}. In view of these costs, DNN models for power-constrained medical wearables should be optimised for energy savings while maintaining prediction accuracy. Model optimisations such as pruning and quantisation can increase sparsity (number of zeroes) and hence greatly reduce model size when combined, providing an opportunity to reduce both the operations performed and volume of data transferred during prediction, for better energy efficiency.

In this work, we develop an optimised convolutional neural network (CNN) that classifies AFib from ECG readings. The CNN model is described in section II, followed by the pruning strategy in section III, and quantization details in section IV. Results and conclusions are presented in sections V and VI respectively.

\section{\label{Related}CNN model for AFib detection}

\subsection{\label{Preprocessing}Preprocessing of ECG Dataset}

The dataset used for the AFib detection in this work is obtained from The PhysioNet Computing in Cardiology Challenge 2017~\cite{clifford2017challenge}, from which ECG lead recordings are classified into four classes: normal sinus rhythm (N), atrial fibrillation (A), other alternate rhythms (O) and noisy readings ($\sim$). The dataset contained a total of 8,528 labelled ECG signals. The raw ECG data was standardised with zero mean and unit variance. Due to the episodic nature of AFib, a fixed input length of 18,000 (close to maximum signal length) was used to avoid losing key information of longer signals. Shorter data were zero-padded to the standard input length, whereas longer signals were trimmed.

\begin{table}[htbp!]
\centering
\caption{Distribution of zero-padded data versus total distribution of classes}
\begin{tabular}{|m{2cm}<{\centering}|m{0.8cm}<{\centering}|m{0.8cm}<{\centering}|m{0.8cm}<{\centering}|m{0.8cm}<{\centering}|m{0.8cm}<{\centering}|}
\hline
                                & A     & N     & O     & $\sim$    & Total \\ \hline
Shorter than 18,000 (padded)    & 662   & 4665  & 2081  & 272       & 7680  \\ \hline
Total in dataset                & 758   & 5076  & 2415  & 279       & 8528  \\ \hline
Distribution of padded data     & 8.62\% & 60.7\% & 27.1\% & 3.54\% & 90.06\%   \\ \hline
Total distribution              & 8.89\% & 59.5\% & 28.3\% & 3.27\% & -     \\ \hline
\end{tabular}
\label{tbl:padded_distribution}
\end{table}

To evaluate the risk of introducing bias towards predicting certain classes through zero-padding, the distribution of padded data by class was compared to the total distribution of classes. As the two distributions were remarkably similar, the padded data was analysed to have minimal risk in skewing the model towards any particular class. Table~\ref{tbl:padded_distribution} shows the distribution of padded data in terms of classes.

\begin{figure}[htbp!]
\centering
\includegraphics[width=\myfigWidth]{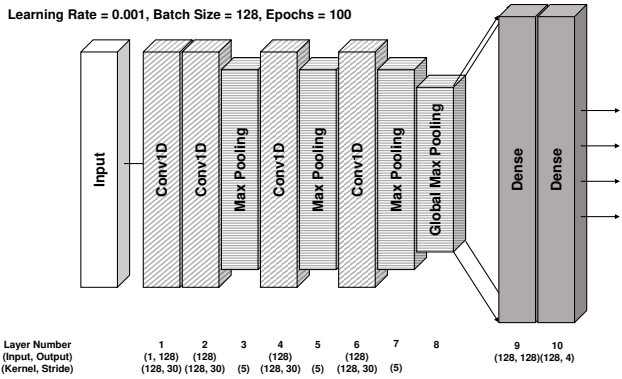}
\caption{Model architecture of CNN model. The model performs one-dimensional convolutions in convolutional layers (signified with slash patterns). Dimensions are denoted as (length of layer, number of channels)}
  \label{fig:base_cnn_model}
\end{figure}

\subsection{\label{Architecture}CNN Model Architecture}

The model architecture designed is pictured in Figure~\ref{fig:base_cnn_model}. The model comprises four convolution layers with 128 channels and two dense layers. Rectified linear unit (ReLU) was used as the activation function for all the layers excluding the final dense layer (which utilises the softmax function for classification). The use of ReLU sets negative outputs to zero, which increases sparsity in output feature maps. This sparsity can be exploited by data compression algorithms to reduce data transfer energy costs and improve efficiency~\cite{chen2016eyeriss}.

Max pooling layers were used to downsize outputs of convolution layers by a factor of 5. Max pooling was intended to select important features within the feature map while also maintaining a smaller model size. The input to the model are long ECG vectors that contains 18,000 samples each, which makes the features extracted by the first convolutional layer less representative. From our experimentation, performing max pooling only after the first two convolutional layers performed better. We deduce that this design allows more filters to be applied and patterns in the input signals are more enhanced and better represented before the inputs are downsized. 

\subsection{\label{Training}Model Training}

\begin{figure}[htbp!]
  \centering
  \subfloat[]{
  \begin{minipage}[b]{\myfigWidth}
  \includegraphics[width=\myfigWidth]{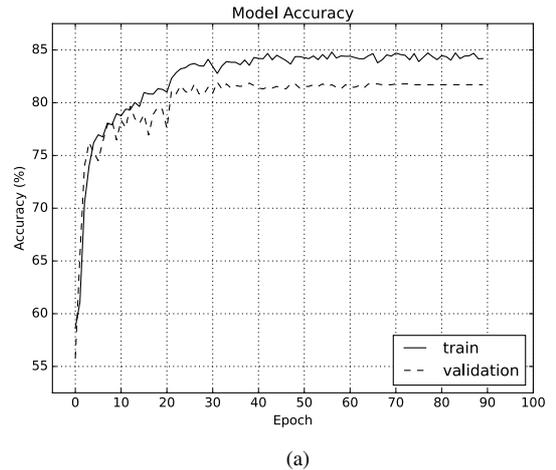}
  \label{fig:train_acc}
  \end{minipage}
  }\\
  \subfloat[]{
   \begin{minipage}[b]{\myfigWidth}
  \includegraphics[width=\myfigWidth]{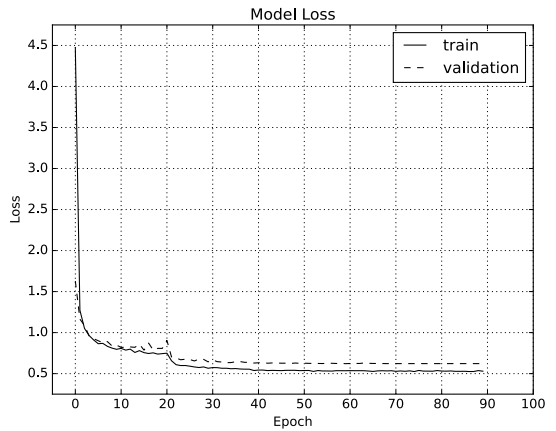}
  \label{fig:train_loss}
  \end{minipage}
  }
  
  \caption{Model performance during training:~\protect\subref{fig:train_acc} model accuracy,~\protect\subref{fig:train_loss} model loss}
  \label{fig:train_eval}
\end{figure}

For training, the dataset was divided into 70\% training set, 15\% validation set, and 15\% testing set to evaluate model performance. The validation set was used to avoid overfitting by evaluating both training loss and validation loss during training. A similar distribution of the four classes across all three sets is enforced. Training was performed with a batch size of 128 and a maximum of 100 epochs. Figure~\ref{fig:train_eval} indicates variation of model accuracy and loss against training epochs. The CNN’s hyperparameters were decided using grid search for hyperparameter tuning~\cite{hparams}. The tuning was done in conjunction with 5-fold cross validation to ensure robustness, as the model was tested 5 times with 5 different sets of testing data. Weight decay was also implemented to prevent overfitting~\cite{sklearn}.

\subsection{\label{Performance_Non_Optimised}Model Performance Before Optimisation}

The trained model achieved a classification accuracy of 92.2\% and a model size (weights and biases) of 16.40 MB. The use of ReLU enabled the model to achieve a high average feature map sparsity of 91.8\% (determined using testing data) even before pruning. However, the trained model had no sparsity in weights and biases as all values were non-zero. Details of the initial model performance and size is tabulated in Table~\ref{tbl:initial_model}.

\begin{table}[htbp!]
\scriptsize
\centering
\caption{Performance and size of initial CNN model}
\begin{tabular}{|m{2.5cm}<{\centering}|m{1.5cm}<{\centering}|}
\hline
Accuracy    & 92.2\%   \\ \hline
Precision   & 84.3\%   \\ \hline
Specificity & 94.8\%   \\ \hline
Sensitivity & 84.3\%   \\ \hline
F1-Score    & 84.3\%   \\ \hline
Model Size (weights and biases)(MB) & 16.40 \\ \hline
Model Sparsity (weights and biases) & 0\%   \\ \hline
\end{tabular}
\label{tbl:initial_model}
\end{table}

\section{\label{Pruning}Model Pruning}

Pruning aims to remove less influential parameters (weights, filters, etc.) by setting these parameters to zero. In doing so, pruning increases the sparsity of a model, potentially reducing computations required for inference and the overall model size~\cite{sze2020sparsity}. However, pruning the network typically results in a drop in accuracy metrics due to the loss of information from the model. Therefore, the objective of pruning is to minimise the loss and drop in accuracy while achieving the highest compression ratio (highest sparsity) possible for the model~\cite{zhang2020iot}. 

Two pruning methods were explored in this study: magnitude-based weight pruning and filter pruning. Magnitude-based weight pruning is a fine-grained pruning method that approximates the sensitivity of weights by magnitude. Smaller weights are removed first, followed by weights with a greater value, until a set degree of sparsity is reached. After each iteration, the network is fine-tuned globally. This method is widely used to optimise and downsize the model quickly~\cite{yang2017energy}.

In contrast, filter pruning is a coarse-grained pruning method that removes entire filters if they do not produce any productive outputs. An algorithm by Li et al.~\cite{li2018filterpruning} uses feature selection by k-means++ for filter pruning. Using the algorithm, filters that perform similar functions (and are hence highly correlated) are consolidated and the redundant filters are removed. An adaptation of the algorithm by Vecsei~\cite{vecsei2019github} was applied to our model to study the compression ratio achieved.

\subsection{\label{Pruning_Result}Results of Model Pruning}

We applied magnitude-based weight pruning to our model. The results after magnitude-based weight pruning are detailed in Table~\ref{tbl:pruned_model}. The sparsity of the model, measured by counting zeroes in weights and biases, increased from 0\% to 90\%. The model size was reduced from 16.40 MB to 1.08 MB, achieving a compression ratio of 15.2×. The overall accuracy metrics were maintained after pruning, with some metrics even slightly improved (such as sensitivity improvement of 0.7\%). The slight improvement could be a result of undesirable weights being removed, which had originally confused the model. There were more than 1.5 million parameters in the original model, which makes it less sensitive to parameter pruning.  In this case, a higher final sparsity level could be set for the model pruning process without significant drop in model performance. However, since the model size at a sparsity level of 90\% was already suitable for hardware implementation, the pruning process was terminated and this model was taken as the final result of the pruning process. The filter pruning algorithm of~\cite{vecsei2019github} did not prune any filters when applied to the model. Hence, there were no changes to the model and no compression was achieved. This suggests that filters in the developed model had little correlation and so was unsuitable for the k-means++ algorithm to be applied. Therefore, the weight-pruned model was used as the final model.

\begin{table}[htbp!]
\scriptsize
\centering
\caption{Performance and size of pruned CNN model}
\begin{tabular}{|m{2.5cm}<{\centering}|m{1.5cm}<{\centering}|}
\hline
Accuracy    & 92.5\%   \\ \hline
Precision   & 85.0\%   \\ \hline
Specificity & 95.0\%   \\ \hline
Sensitivity & 85.0\%   \\ \hline
F1-Score    & 85.0\%   \\ \hline
Model Size (weights and biases)(MB) & 1.08 \\ \hline
Model Sparsity (weights and biases) & 90\%   \\ \hline
\end{tabular}
\label{tbl:pruned_model}
\end{table}

\section{\label{Quantisation}Logarithmic Quantisation}

Quantisation involves reducing the precision of feature maps and/or filter weights to reduce the size of model data transferred~\cite{guoxin_icta20}. Quantisation can improve energy efficiency of DNN accelerator systems, as energy costs of computation scale with data bit-width~\cite{yang2017energy}. Furthermore, combining quantisation with other optimisations such as pruning can further reduce model size as well. In this work, logarithmic quantisation is employed. Logarithmic quantisation operates on the notion that non-uniformly distributed weights can be approximated as base-2 logarithmic representations. By limiting the range of weight values to powers of two, only exponent values of weights need to be stored and transferred. These exponents can be represented as integers with much fewer bits. For example, if weight values are limited to exponents of $-7$ to $7$ (meaning the values $2^{-7}$, $2^{-6}$, …, $2^{7}$), each weight can be represented by only a 4-bit signed integer value. Furthermore, logarithmic quantisation can also reduce the computation power required for convolution, as multiplication of weights with feature maps can use only shifters for bit-shifting~\cite{miyashita2016logarithmic}.

We implemented logarithmic quantisation with reference to~\cite{miyashita2016logarithmic} alongside a rounding function as described in~\cite{cai2018logarithmic}. Weight values were limited to $\pm2^{-E}$, in which $E$ is the magnitude of the exponent. Hence, weight values can be stored as the unsigned integer $E$ and an additional bit for the weight’s sign. For example, $-2^{-3}$ is stored as $-3$ and $+2^{-4}$ is stored as $+4$. Two ranges of exponent magnitude $E$ were explored in this study: 2-bit magnitudes ($0$ to $3$) and 3-bit magnitudes ($0$ to $7$). In the example of a 3-bit magnitude, the weight is stored using 4 bits, with one bit for the weight’s sign. Weights were constrained to these values before training. Logarithmically quantised weights that are too low in value are treated as having minimal contribution and are clipped to $0$, much like magnitude-based pruning.

\subsection{\label{Quantisation_Result}Results of Logarithmic Quantisation}

Logarithmic quantisation using the 3-bit magnitudes demonstrated favourable results, with only a 0.3\% decrease in classification accuracy and less than 0.6\% loss across all metrics, comparing to the results obtained with the original model in Table~\ref{tbl:initial_model}. Model size was reduced from 16.40 MB to 11.30 MB with weights still represented in floating point. Converting the weights and biases to 4-bit values (3-bit magnitude, 1-bit sign) yielded a smaller data size of 0.74 MB, as is shown in Table~\ref{tbl:quantised_model}.

However, using 2-bit magnitudes resulted in a drastic drop in accuracy to 52.1\% and drop in sensitivity to 4.2\%. The model had predicted all data as noisy ($\sim$ class). Hence, 2-bit magnitudes for exponents were considered too aggressive as it would severely limit the range of actual weight values. Sparsity introduced was comparable to that of 3-bit magnitudes, at 33.4\%. Results for both magnitude ranges are tabulated in Table~\ref{tbl:quantised_model}. Table~\ref{tbl:quant_eval} shows results obtained in separate runs at greater magnitude ranges. The model hardly benefited from bit-width increment with magnitude range greater than 3-bit. A 3-bit multiplier which has a structure that has been evolved by Vassilev contains 37 total LUTs~\cite{larchev2004hardware}, while a 16-bit floating point RNS multiplier requires 150 logic elements~\cite{samhitha2013implementation}. 3-bit magnitudes for exponents in the quantisation process could maintain good performance with less hardware resource needed. Hence, 3-bit magnitudes for exponents suited most for logarithmic quantisation.

\begin{table}[htbp!]
\scriptsize
\centering
\caption{Performance and size of quantised CNN model}
\begin{tabular}{|m{3cm}<{\centering}|m{1.5cm}<{\centering}|m{1.5cm}<{\centering}|}
\hline
            & \textbf{3-bit $E$} & \textbf{2-bit $E$} \\ \hline
Accuracy    & 91.9\%             & 52.1\% \\ \hline
Precision   & 83.8\%             & 4.2\%  \\ \hline
Specificity & 94.6\%             & 68.1\% \\ \hline
Sensitivity & 83.8\%             & 4.2\%  \\ \hline
F1-Score    & 83.8\%             & 4.2\%  \\ \hline
Model Size (weights and biases in floating point)(MB) & 11.30 & 10.92 \\ \hline
Model Size (weights and biases in 4-bit values)(MB)   & 0.74  & 0.56  \\ \hline
\end{tabular}
\label{tbl:quantised_model}
\end{table}

\begin{table}[htbp!]
\scriptsize
\centering
\caption{Performance and size of quantised CNN model}
\begin{tabular}{|m{1cm}<{\centering}|m{0.95cm}<{\centering}|m{0.95cm}<{\centering}|m{0.95cm}<{\centering}|m{0.95cm}<{\centering}|m{0.95cm}<{\centering}|}
\hline
            & \textbf{2-bit $E$} & \textbf{3-bit $E$} & \textbf{4-bit $E$} & \textbf{8-bit $E$} & \textbf{16-bit $E$} \\
\hline
Accuracy    & 51.5\% & 89.2\% & 89.4\% & 89.4\% & 89.4\% \\
\hline
\end{tabular}
\label{tbl:quant_eval}
\end{table}

\section{\label{Combined_Result}Combined Optimisation Results}

Combining both magnitude-based weight pruning and logarithmic quantisation with 3-bit magnitudes on the model, the optimisations achieved significant size reduction from 16.40 MB to 0.18 MB. Classification accuracy was at 91.7\%, demonstrating minimal loss of less than 1\% across all metrics. The combined results are tabulated in Table~\ref{tbl:comparing_model}.

\begin{table}[htbp!]
\scriptsize
\centering
\caption{Final results of optimised CNN model}
\begin{tabular}{|m{3cm}<{\centering}|m{1.5cm}<{\centering}|m{1.5cm}<{\centering}|}
\hline
            & \textbf{Original Model} & \textbf{Optimised Model} \\ \hline
Accuracy    & 92.2\%             & 91.7\% \\ \hline
Precision   & 84.3\%             & 83.4\%  \\ \hline
Specificity & 94.8\%             & 94.5\% \\ \hline
Sensitivity & 84.3\%             & 83.4\%  \\ \hline
F1-Score    & 84.3\%             & 83.4\%  \\ \hline
\textbf{Model Size (weights and biases)(MB)} & \textbf{16.40} & \textbf{0.18} \\ \hline
\end{tabular}
\label{tbl:comparing_model}
\end{table}

Other model structures are compared with our optimised model in terms of overall F1 score. Tziakouri et al.~\cite{tziakouri2017classification} trained two Support Vector Machines (SVM) with statistical features extracted from ECGs and achieved an overall F1 score of 0.66. In~\cite{jimenez2017atrial}, a Feedforward Neural Network (FFNN) was implemented and trained with selected features obtained automatically from ECGs, and reached an overall F1 score of 0.77 with the final tuned model. Chandra et al.~\cite{chandra2017atrial} fed the CNN with short ECG segments that contains only 3 R-peaks, and achieved an overall F1 score at 0.71 with a light-weight model. In~\cite{teijeiro2017arrhythmia}, Recurrent Neural Network (RNN) method was adopted and Long Short Term Memory networks (LSTM) was implemented to exploit the nature of LSTM that they are able to remember information of a long sequential type of data using a cell state. The RNN method obtained an overall F1 score of 0.83. As a comparison, our optimised model reached an overall F1 score of 0.84, which is the best result among the methods tabulated in Table~\ref{tbl:comparing_method}.

\begin{table}[htbp!]
\centering
\caption{Overall F1 score comparing with other methods}
\begin{tabular}{|m{4cm}<{\centering}|m{2cm}<{\centering}|}
\hline
\textbf{Method} & \textbf{Overall F1 Score}  \\ \hline
SVM~\cite{tziakouri2017classification} & 0.66 \\ \hline
FFNN~\cite{jimenez2017atrial}          & 0.77 \\ \hline
CNN with R-peak Detection~\cite{chandra2017atrial} & 0.71 \\ \hline
RNN~\cite{teijeiro2017arrhythmia}      & 0.83 \\ \hline
\textbf{Optimised Model}         & \textbf{0.84} \\ \hline
\end{tabular}
\label{tbl:comparing_method}
\end{table}

\section{\label{Conclusion}Conclusion}

A convolutional neural network model was developed to detect atrial fibrillation from ECG signals. The model was trained using a deep learning approach rather than feature-based extraction. A classification accuracy of 92.2\%, and a 91.8\% sparsity in internal feature maps before pruning was achieved. Magnitude-based weight pruning, and logarithmic quantisation were then applied, reducing model size from 16.40 MB to 0.18 MB for a compression ratio of 91.1× with minimal loss of less than 1\% across all metrics. The optimised model developed presents opportunities for reducing energy consumption of a hardware implementation when sparsity and reduced precision is appropriately exploited. Achieving energy efficiency while maintaining high prediction accuracy will be essential for implementing DNNs on battery-powered medical devices.




%




\bibliographystyle{IEEEtran}
\bibliography{EMBC.bib}

\begin{thebibliography}{10}
\providecommand{\url}[1]{#1}
\csname url@samestyle\endcsname
\providecommand{\newblock}{\relax}
\providecommand{\bibinfo}[2]{#2}
\providecommand{\BIBentrySTDinterwordspacing}{\spaceskip=0pt\relax}
\providecommand{\BIBentryALTinterwordstretchfactor}{4}
\providecommand{\BIBentryALTinterwordspacing}{\spaceskip=\fontdimen2\font plus
\BIBentryALTinterwordstretchfactor\fontdimen3\font minus
  \fontdimen4\font\relax}
\providecommand{\BIBforeignlanguage}[2]{{%
\expandafter\ifx\csname l@#1\endcsname\relax
\typeout{** WARNING: IEEEtran.bst: No hyphenation pattern has been}%
\typeout{** loaded for the language `#1'. Using the pattern for}%
\typeout{** the default language instead.}%
\else
\language=\csname l@#1\endcsname
\fi
#2}}
\providecommand{\BIBdecl}{\relax}
\BIBdecl

\bibitem{hannun2019cardiologist}
A.~Y. Hannun, P.~Rajpurkar, M.~Haghpanahi, G.~H. Tison, C.~Bourn, M.~P.
  Turakhia, and A.~Y. Ng, ``Cardiologist-level arrhythmia detection and
  classification in ambulatory electrocardiograms using a deep neural
  network,'' \emph{Nature medicine}, vol.~25, no.~1, pp. 65--69, 2019.

\bibitem{xiaolin_icecs20}
L.~Xiaolin, B.~Cardiff, and D.~John, ``A {1D} convolutional neural network for
  heartbeat classification from single lead {ECG},'' in \emph{2020 27th IEEE
  International Conference on Electronics, Circuits and Systems (ICECS)}, 2020,
  pp. 1--2.

\bibitem{maryam_tbiocas2021}
M.~Saeed, Q.~Wang, O.~Martens, B.~Larras, A.~Frappe, B.~Cardiff, and J.~Deepu,
  ``Evaluation of level-crossing adcs for event-driven ecg classification,''
  \emph{IEEE Transactions on Biomedical Circuits and Systems}, pp. 1--1, 2021,
  doi: \url{10.1109/TBCAS.2021.3136206}.

\bibitem{ANNet}
G.~Sivapalan, K.~Nundy, S.~Dev, B.~Cardiff, and J.~Deepu, ``Annet: A
  lightweight neural network for ecg anomaly detection in iot edge sensors,''
  \emph{IEEE Transactions on Biomedical Circuits and Systems}, pp. 1--1, 2022,
  doi: \url{10.1109/TBCAS.2021.3137646}.

\bibitem{colloca2013mobile}
R.~Colloca, ``Implementation and testing of a trial fibrillation detectors for
  a mobile phone application,'' Master's thesis, Politecnicodi Milano and
  University of Oxford, 2013.

\bibitem{zhang2020iot}
Z.~Zhang and A.~Z. Kouzani, ``Implementation of {DNN}s on {IoT} devices,''
  \emph{Neural Computing and Applications}, vol.~32, no.~5, pp. 1327--1356,
  2020.

\bibitem{chen2016eyeriss}
Y.-H. Chen \emph{et~al.}, ``Eyeriss: An energy-efficient reconfigurable
  accelerator for deep convolutional neural networks,'' \emph{IEEE journal of
  solid-state circuits}, vol.~52, no.~1, pp. 127--138, 2016.

\bibitem{clifford2017challenge}
G.~D. Clifford, C.~Liu, B.~Moody, H.~L. Li-wei, I.~Silva, Q.~Li, A.~Johnson,
  and R.~G. Mark, ``{AF} classification from a short single lead {ECG}
  recording: The physionet/computing in cardiology challenge 2017,'' in
  \emph{2017 Computing in Cardiology (CinC)}.\hskip 1em plus 0.5em minus
  0.4em\relax IEEE, 2017, pp. 1--4.

\bibitem{hparams}
TensorFlow, ``Hyperparameter tuning with the {HP}arams dashboard,''
  \url{https://www.tensorflow.org/tensorboard/hyperparameter\_tuning\_with\_\\hparams},
  accessed: 2020-10-26.

\bibitem{sklearn}
scikit learn, ``sklearn.model\_selection.kfold,''
  \url{https://scikit-learn.org/stable/modules/generated/sklearn.model\_selection.KFold.html},
  accessed: 2021-02-21.

\bibitem{sze2020sparsity}
T.~J.~Y. V.~Sze, Y. H.~Chen and J.~S. Emer, ``Exploiting sparsity,'' in
  \emph{Efficient Processing of Deep Neural Networks}.\hskip 1em plus 0.5em
  minus 0.4em\relax California: Morgan \& Claypool Publishers, 2020, pp.
  167--228.

\bibitem{yang2017energy}
T.-J. Yang, Y.-H. Chen, and V.~Sze, ``Designing energy-efficient convolutional
  neural networks using energy-aware pruning,'' in \emph{Proceedings of the
  IEEE Conference on Computer Vision and Pattern Recognition}, 2017, pp.
  5687--5695.

\bibitem{li2018filterpruning}
L.~Li, Y.~Xu, and J.~Zhu, ``Filter level pruning based on similar feature
  extraction for convolutional neural networks,'' \emph{IEICE TRANSACTIONS on
  Information and Systems}, vol. 101, no.~4, pp. 1203--1206, 2018.

\bibitem{vecsei2019github}
G.~Vecsei, ``Filter pruning in deep convolutional networks,''
  \url{https://github.com/gaborvecsei/Ridurre-Network-Filter-Pruning-Keras/tree/v0.0.2},
  2019, accessed: 2021-02-22.

\bibitem{guoxin_icta20}
G.~Wang, D.~John, and A.~Nag, ``Low complexity {ECG} biometric authentication
  for {IoT} edge devices,'' in \emph{2020 IEEE International Conference on
  Integrated Circuits, Technologies and Applications (ICTA)}, 2020, pp.
  145--146.

\bibitem{miyashita2016logarithmic}
D.~Miyashita, E.~H. Lee, and B.~Murmann, ``Convolutional neural networks using
  logarithmic data representation,'' \emph{arXiv preprint arXiv:1603.01025},
  2016.

\bibitem{cai2018logarithmic}
J.~Cai \emph{et~al.}, ``A deep look into logarithmic quantization of model
  parameters in neural networks,'' in \emph{Proceedings of the 10th
  International Conference on Advances in Information Technology}, 2018, pp.
  1--8.

\bibitem{larchev2004hardware}
G.~V. Larchev and J.~D. Lohn, ``Hardware-in-the-loop evolution of a 3-bit
  multiplier,'' in \emph{12th Annual IEEE Symposium on Field-Programmable
  Custom Computing Machines}.\hskip 1em plus 0.5em minus 0.4em\relax IEEE,
  2004, pp. 277--278.

\bibitem{samhitha2013implementation}
N.~R. Samhitha \emph{et~al.}, ``Implementation of 16-bit floating point
  multiplier using residue number system,'' in \emph{2013 International
  Conference on Green Computing, Communication and Conservation of Energy
  (ICGCE)}.\hskip 1em plus 0.5em minus 0.4em\relax IEEE, 2013, pp. 195--198.

\bibitem{tziakouri2017classification}
M.~Tziakouri \emph{et~al.}, ``Classification of {AF} and other arrhythmias from
  a short segment of ecg using dynamic time warping,'' in \emph{2017 Computing
  in Cardiology (CinC)}.\hskip 1em plus 0.5em minus 0.4em\relax IEEE, 2017, pp.
  1--4.

\bibitem{jimenez2017atrial}
S.~Jim{\'e}nez-Serrano, J.~Yag{\"u}e-Mayans, E.~Simarro-Mond{\'e}jar, C.~J.
  Calvo, F.~Castells, and J.~Millet, ``Atrial fibrillation detection using
  feedforward neural networks and automatically extracted signal features,'' in
  \emph{2017 Computing in Cardiology (CinC)}.\hskip 1em plus 0.5em minus
  0.4em\relax IEEE, 2017, pp. 1--4.

\bibitem{chandra2017atrial}
B.~Chandra, C.~S. Sastry, S.~Jana, and S.~Patidar, ``Atrial fibrillation
  detection using convolutional neural networks,'' in \emph{2017 Computing in
  Cardiology (CinC)}.\hskip 1em plus 0.5em minus 0.4em\relax IEEE, 2017, pp.
  1--4.

\bibitem{teijeiro2017arrhythmia}
T.~Teijeiro, C.~A. Garc{\'\i}a, D.~Castro, and P.~F{\'e}lix, ``Arrhythmia
  classification from the abductive interpretation of short single-lead {ECG}
  records,'' in \emph{2017 Computing in cardiology (cinc)}.\hskip 1em plus
  0.5em minus 0.4em\relax IEEE, 2017, pp. 1--4.

\end{thebibliography}
\end{document}